\documentclass[%
 preprint,
 amsmath,amssymb,
 prd,
]{revtex4-1}
\usepackage{hyperref}
\usepackage{graphicx}
\usepackage{caption}
\usepackage{subcaption}
\usepackage{dcolumn}
\usepackage{bm}

\begin{document}

\title{On analytical solutions of the DIS structure functions in the framework of GLR-MQ-ZRS equation  }

\author{Madhurjya Lalung}
\email{mlalung2016@gmail.com(Corresponding author)}
\author{Pragyan Phukan}
\email{pragyanp@tezu.ernet.in}
\author{Jayanta K. Sarma}
\email{jks@tezu.ernet.in}
\affiliation{
 HEP Laboratory, Department of Physics, Tezpur University, Tezpur 784028, Assam, India }

\date{\today}

\begin{abstract}
We computed the deep inelastic scattering (DIS) structure functions $F_2 (x,Q^2)$ and $F_L (x,Q^2)$ in the framework of Gribov-Levin-Ryskin-Mueller-Qiu, Zhu-Ruan-Shen (GLR-MQ-ZRS) equation. Both $x$ and $Q^2$ evolutions of the structure functions are studied in the kinematic range of $10^{-6}\leq x \leq 10^{-2}$ and $5 \leq Q^2 \leq 100\,,GeV^2$ respectively. we have suggested analytical solutions to the structure functions, our results are inspired by Regge like behaviour of gluons at small-x. The solutions of structure functions are obtained for two particular cases: (i) treating the strong coupling constant $\alpha_s$ as an arbitrary constant, and (ii) considering the $Q^2$ dependency of $\alpha_s$. Our predicted results are in good agreement with the latest available high precision HERA data. 
  
\end{abstract}

\maketitle


\section{Introduction}
\label{intro}
In the phenomenology of Quantum Chromodynamics (QCD), Dokshitzer-Gribov-Lipatov-Altarelli-Parisi  (DGLAP) evolution equations\cite{dokshitzer77,altarelli1977asymptotic,gribov72} serve as basic tools to study the $x$ and $\ln Q^2$ evolutions of structure functions, where $x$ and $Q^2$ are the momentum fraction of proton carried by the struck parton and the four momentum transfer squared of deep inelastic scattering process respectively (DIS). DIS
processes are important tools also to explore the internal structure of nucleons in high-energy scattering at the CERN Large
Hadron Collider, Fermilab Tevatron collider, and in other
hard scattering experiments. The measurements of the $F_2$ and $F_L$ structure functions by the DIS has opened up new era in measurements of the parton densities inside hadrons. These structure functions reflect the momentum distributions of partons in a nucleon and are directly related
to parton distribution functions (PDFs).  At small-x values, since gluons dominate among the partons therefore, the dominant contribution to $F_2 (x,Q^2)$ as well as $F_L(x, Q^2)$ comes from the gluon operators. Hence measurement of  $F_2 (x,Q^2)$ and  $F_L (x,Q^2)$ can be used in order to extract the gluon distribution functions and therefore, the measurement these structure functions become instrumental in providing sensitive test of perturbative QCD. The most precise measurements on the  inclusive structure function has been provided by the electron-proton collider at HERA which has brought remarkable improvements to our understanding on the structure of proton and can be used to extract high precision
PDFs. Many high precision measurements of the cross sections or other observables at the colliders rely on high precision PDFs for any processes that involve
colliding of hadrons.

In the moderate kinematical region of $x\,(x \geq  10^{-2})$, the well established DGLAP equations,
which are based upon the sum of QCD ladder diagrams have been successfully applied to describe the available HERA data\cite{khanpour2017analytic,khanpour2013new,ZARRIN2017126,Boroun2014}. HERA data shows\cite{Andreev2,ADLOFF1997452,ADLOFF19973,Adloffetal.2001} that the gluon distribution
function has a steep behaviour in this region. However, at small x, the problem becomes more complicated since this strong
growth of gluons can eventually lead to violation of unitarity and Froissart bound \cite{froissart1961asymptotic} on the physical cross-sections. So, this undesirable growth has to be
tamed by screening effects. These screening effects are provided by gluon recombinations in the high dense gluon regime, multiple gluon interactions lead to nonlinear terms in the DGLAP equation. 
These nonlinear terms are responsible for slowing down the unusual growth of the gluon densities in this kinematic region where $\alpha_s$ is still small but
the density of partons becomes large. Gribov, Levin,
Ryskin, Mueller and Qiu (GLR-MQ)\cite{GRIBOV19831,MUELLER1986427} carried out
detailed study of this region and argued that the
physical processes of parton interactions and parton recombinations become important in the parton cascade when the parton density is large, and these
shadowing corrections could be expressed in a new
evolution equation known as the GLR-MQ equation\cite{GRIBOV19831,MUELLER1986427}. 

However, certain issues of the GLR-MQ equation cannot be overlooked. Firstly, the application of the AGK(Abramovsky, Gribov,Kanchelli) cutting rule in the GLR-MQ corrections breaks important evolution kernels\cite{zhu}; Secondly, momentum conservation is violated in the GLR-MQ equation\cite{zhu1993antishadowing}; and thirdly, since the GLR-MQ equation was derived at Double Leading Logarithmic approximation(DLLA) which is valid at small-x,  the nonlinear corrections cannot smoothly connect with the DGLAP equation\cite{zhu1}. These problems motivated Zhu and his co authors (Ruan, Shen) to rederive DGLAP equation by including recombination probabilities of partons in the leading logarithmic approximation (LL($Q^2$)A) in a time ordered perturbation theory(TOPT) framework, resulting in a new evolution equation (the GLR-MQ-ZRS equation)\cite{zhu,zhu1993antishadowing,zhu1}. Unlike the GLR-MQ equation, momentum conservation is restored naturally in GLR-MQ-ZRS equation. The main difference between GLR-MQ equation and ZRS version is the presence of an additional antishadowing term in the latter. In our previous work\cite{LALUNG2019121615}, we have obtained solutions for gluon distribution function $G(x,Q^2)$ and we had shown that our predicted results of proton's $F_2$ derivative $dF_2 /\ln Q^2$ were in good agreement with the HERA data. In this work, however, we have obtained the solution of $F_2 (x,Q^2)$ and $F_L (x,Q^2)$ structure functions from the GLR-MQ-ZRS equation. Our predicted results are compared with the high precision HERA data. 

The paper is organised as follows: In section \ref{sec2}, we develop the formalism to obtain solutions of $F_2 (x,Q^2)$ and $F_L (x,Q^2)$ from GLR-MQ-ZRS equation. The solutions are obtained by considering both $\alpha_s$ fixed and $\alpha_s$ running. We discuss about our results in section \ref{sec3}, and we conclude in section \ref{sec4}.

\section{Formalism}\label{sec2}
\label{formalism}
In the parton model framework, the structure functions are usually identified by summing quark distributions weighted by their respective charges squared 

\begin{equation}\label{1}
F_2(x, Q^2) = \sum_{i} e_i ^2 xq_i(x, Q^2)
\end{equation}
where the sum implies summation over all flavours of quarks and anti-quarks, $e_i$ is the charge of quark i. The $F_2$ structure function measured in DIS experiments has contributions coming from both the singlet and nonsinglet terms. But, at small-x, the nonsinglet contributions are negligible. Therefore, Eq. \ref{1} can be effectively written as  

\begin{equation}\label{2}
F_2 (x,Q^2) \approx \frac{5}{18}F_2 ^S (x,Q^2)
\end{equation} 
At small-x, the density of gluons are so high that they dominate over all the partons. In this work, we will explore this region where the density of gluons are so high and the recombination effects become important. The GLR-MQ-ZRS equation for the singlet structure function is given by

\begin{equation}\label{3}
\begin{split}
\frac{d F_2 ^S (x, Q^2)}{d \ln Q^2} = &\frac{2T_f \alpha_s}{\pi}\int_{x}^{1}\frac{dx_1}{x_1}\frac{x}{x_1}P_{AP} ^{G\rightarrow q} (x_1,x)x_1 G(x_1,Q^2)\\& + \frac{\alpha_s ^2}{4\pi R^2 Q^2}\int_{x/2}^{1/2}dx_1 x{x_1}G^2(x_1,Q^2)\sum_{i} P_{i} ^{GG\rightarrow q} (x_1,x)\\&
-\frac{\alpha_s ^2}{4\pi R^2 Q^2}\int_{x}^{1/2}dx_1 x{x_1}G^2(x_1,Q^2)\sum_{i} P_{i} ^{GG\rightarrow q} (x_1,x),
\end{split}
\end{equation}

where $P_{AP} ^{G\rightarrow q}$ are the AP splitting kernels\cite{furmanski1980singlet}, $\alpha_s$ is the strong coupling constant, $R$ is the correlative radius of the gluons inside the hadrons. The typical value of $R$ is considered to be $5\, GeV^{-1}$ which is the size of the proton. If the gluons are concentrated at hotspots then $R$ can be considered $\approx2\,GeV^{-1}$. $T_f = n_f/2$, $n_f$ is the number of flavours. The recombination probabilities are given by
$$P_{i} ^{GG\rightarrow q} (x_1,x)= \frac{1}{96}\cdot\frac{(2x_1-x)^2(8x_1 ^2-21x_1 x+ 14x ^2)}{x_1 ^5}$$

The first term of Eq. \ref{3} is the usual DGLAP term. The second and third term represents the positive antishadowing and shadowing term respectively. Both these terms have separate kinematical regions. Eq. \ref{3} can be read as

\begin{equation}\label{4}
\begin{split}
\frac{d F^S _2(x, Q^2)}{d \ln Q^2} = \frac{5}{18}\cdot\frac{d F^S _2(x, Q^2)}{d \ln Q^2}\bigg|_{DGLAP} &+ \frac{\alpha_s ^2}{4\pi R^2 Q^2}\int_{x/2}^{1/2}dx_1 x{x_1}G^2(x_1,Q^2)\sum_{i} P_{i} ^{GG\rightarrow q} (x_1,x)\\
&-\frac{\alpha_s ^2}{4\pi R^2 Q^2}\int_{x}^{1/2}dx_1 x{x_1}G^2(x_1,Q^2)\sum_{i} P_{i} ^{GG\rightarrow q} (x_1,x),
\end{split}
\end{equation}

Since, at small-x gluons form high dense system inside the hadrons, the singlet structure functions become sensitive to the gluon distribution function.  It is,
therefore, reasonable to have a direct relationship between gluon distribution
$G(x, Q^2 )$ and the proton structure function
$F_2 (x, Q^2)$, to facilitate direct extraction of $G(x,Q^2)$ from the data of $F_2 (x,Q^2)$. This expectation has led to an approximate
phenomenological scheme, as reported by the authors\cite{SARMA1993323,SARMA1997139,baishya2009evolution,Devee2012} in the past two decades an ansatz relating the gluon
distribution function to singlet structure function. The
commonly used relation is
\begin{equation}\label{5}
G(x, Q^2 ) = K(x)F_2 ^S (x, Q^2),
\end{equation}

where $K(x)$ is a parameter to be chosen from the
experimental data. The actual functional form of $K(x)$ can be determined by simultaneous solutions of coupled equations of singlet structure functions and gluon parton densities which is very difficult to perform,  nevertheless it is beyond the scope of this paper. Previously, the authors have performed phenomenological analysis by considering the function $K(x)$ as an arbitrary constant parameter $K$ for a particular range of x and $Q^2$ under study.  But, in this work we will assume a simple functional form $K(x)= ax^b$, the parameters $a$ and $b$ will be determined by analysing the available experimental data for the proton's structure function. Using the ansatz of Eq. \ref{5}, the Eq. \ref{4} can be re-written as

\begin{equation}\label{6}
\begin{split}
\frac{d F^S _2(x, Q^2)}{d \ln Q^2} = \frac{5}{18}\cdot\frac{d F^S _2(x, Q^2)}{d \ln Q^2} &+ \frac{\alpha_s ^2 K^2 (x)}{4\pi R^2 Q^2}\int_{x/2}^{1/2}dx_1 x{x_1}[F_2^S(x_1,Q^2)]^2\sum_{i} P_{i} ^{GG\rightarrow q} (x_1,x)\\
&-\frac{\alpha_s ^2 K^2 (x)}{4\pi R^2 Q^2}\int_{x}^{1/2}dx_1 x{x_1}[F_2^S(x_1,Q^2)]^2\sum_{i} P_{i} ^{GG\rightarrow q} (x_1,x),
\end{split}
\end{equation}

The DGLAP equation for the singlet structure function at DLLA is given by

\begin{equation}\label{7}
\begin{split}
\frac{d F^S _2(x, Q^2)}{d \ln Q^2}\bigg|_{DGLAP} =& \frac{5}{18}\cdot \frac{\alpha_s}{3\pi}\bigg[ 3 + 4\ln (1-x)F_2 ^S (x,Q^2)\\+ &\int_{x}^{1}\frac{2}{1-x_1}\left( (1+x_1 ^2)F_2 ^S (x/x_1,Q^2) - 2 F_2 ^S (x,Q^2)  \right)dx_1 \\&+n_f K(x)\int_{x}^{1}\left( x_1 ^2 + (1-x_1)^2\right)F_2 ^S(x/x_1,Q^2)dx_1  \bigg],
\end{split}
\end{equation}

It is also interesting to note that in the analysis of HERA experimental data, the behaviour of structure functions towards small-x can be described successfully in the framework of Regge theory\cite{collins1977introduction}. Regge theory provides naive and frugal parametrization of all total cross sections for a wide range of $Q^2$ at very small values of $x<0.07$ \cite{DONNACHIE1992227,DONNACHIE1998408}. The high energy behaviour of hadronic cross sections particularly have two contributions: (i) exchange of pomeron proliferating rise of $F_2 (x,Q^2)$ towards small-x, (ii) exchange of reggeon which is related to the meson trajectories\cite{DONNACHIE1992227,DONNACHIE1998408}. In this Regge pole model the DIS structure function $F_2(x, Q^2)$ are parametrized at small-x as $F_2 \propto x^{-\lambda}$ with $\lambda >0$ being a constant called the Regge intercept. Regge theory also states that the small-x behaviour of both the sea quarks as well as gluons are governed by the same singularity factor in the complex plane of angular momentum\cite{collins1977introduction} and so the same power is expected for the sea quarks and gluons. Moreover, the value of the Regge intercept should be close to 0.5 for all the spin-independent singlet, non-singlet and gluon structure functions applicable in a quite broad range of small-x\cite{soffer1997neutron}. A Regge intercept of 0.5 corresponds to the hard pomeron exchange while soft pomeron exchange has a Regge intercept 0. On these accounts we will employ Regge like ansatz of the singlet structure function $F_2 ^S (x,Q^2)$ to solve Eq. \ref{6}. Thus we assume a simple form of $F_2 ^S (x,Q^2)$ as follows: 

\begin{equation}\label{8}
F_2 ^S (x,Q^2)= f(Q^2)x^{-\lambda}
\end{equation}
After applying Regge ansatz, we recast Eq. \ref{7} and Eq. \ref{6} in terms of the Regge intercept $\lambda$ as 

\begin{equation}\label{9}
\begin{split}
\frac{d F^S _2(x, Q^2)}{d \ln Q^2}\bigg|_{DGLAP} =&  \frac{5\alpha_s}{54\pi}\cdot\bigg[ 3 + 4\ln (1-x)F_2 ^S (x,Q^2)\\&+ \int_{x}^{1}\frac{2}{1-x_1}\left( (1+x_1 ^2)x_1 ^{\lambda} F_2 ^S (x,Q^2) - 2 F_2 ^S (x,Q^2)  \right)dx_1 \\&+n_f K(x)\int_{x}^{1}\left( x_1 ^2 + (1-x_1)^2\right)x_1 ^{\lambda}dx_1 F_2 ^S(x,Q^2)  \bigg]\\
=& \frac{5\alpha_s}{54\pi}\cdot \bigg[  3 + 4\ln (1-x) + \int_{x}^{1}\frac{2(1+x_1 ^2)x_1^{\lambda}-2}{1-x_1}dx_1\\& +n_f K(x)\int_{x}^{1}\left( x_1 ^2 + (1-x_1)^2\right)x_1 ^{\lambda}dx_1   \bigg]F^S _2(x, Q^2)\\
=& \alpha_s \cdot f(x,K)F^S _2(x, Q^2)
\end{split}
\end{equation}

and
\begin{equation}\label{10}
\begin{split}
\frac{d F^S _2(x, Q^2)}{d \ln Q^2} = \alpha_s \cdot f(x,K)F^S _2(x, Q^2)+& \frac{\alpha_s ^2 K^2 (x)}{4\pi R^2 Q^2}\int_{x/2}^{1/2}dx_1 x{x_1}[F_2^S(x_1,Q^2)]^2\sum_{i} P_{i} ^{GG\rightarrow q} (x_1,x) \\
&-\frac{\alpha_s ^2 K^2 (x)}{4\pi R^2 Q^2}\int_{x}^{1/2}dx_1 x{x_1}[F_2^S(x_1,Q^2)]^2\sum_{i} P_{i} ^{GG\rightarrow q} (x_1,x),
\end{split}
\end{equation}

where \begin{equation*}
f(x,K)=\frac{5}{36\pi}\cdot \bigg[  3 + 4\ln (1-x) + \int_{x}^{1}\frac{2(1+x_1 ^2)x_1^{\lambda}-2}{1-x_1}dx_1 +n_f K(x)\int_{x}^{1}\left( x_1 ^2 + (1-x_1)^2\right)x_1 ^{\lambda}dx_1   \bigg]
\end{equation*}

\noindent Now, our main motive is to find out the solutions of Eq. \ref{10} which will give us the evolution of singlet structure function in terms of the Bjorken variable $x$ and resolution scale $Q^2$. We will consider two different cases: (i) $\alpha_s$ will be taken as a constant, and (ii) $\alpha_s$ will be considered as running coupling constant as per QCD. Thus, we proceed as follows:

\subsection{$\alpha_s$ as a constant}

\noindent The GLR-MQ-ZRS equation for the singlet structure function simply becomes

\begin{equation}\label{11}
\begin{split}
\frac{d F^S _2(x, Q^2)}{d \ln Q^2} = \alpha_s \cdot f(x,K)F^S _2(x, Q^2) -\frac{\alpha_s ^2 K^2 (x)}{384 \pi R^2 Q^2}\left( 2^{2\lambda +3} -1 \right)x^2 \xi(\lambda)[F_2^S(x,Q^2)]^2,
\end{split}
\end{equation}
Now, in order to solve Eq. \ref{11}, it is always useful to change the variable $Q^2$ suitably into a new variable $t(=\ln (Q^2/\Lambda ^2),\, \Lambda \,\text{being the QCD cutoff parameter})$. After finding the solution we can easily return back to the original variable $Q^2$. So, Eq. \ref{11} in terms of variable $t$ is re-written as

\begin{equation}\label{12}
\begin{split}
\frac{d F^S _2(x, t)}{d t} = \alpha_s \cdot f(x,K)F^S _2(x, t) -\frac{\alpha_s ^2 K^2 (x)}{384 \pi R^2 \Lambda ^2 e^t}\left( 2^{2\lambda +3} -1 \right)x^2 \xi(\lambda)[F_2^S(x,t)]^2,
\end{split}
\end{equation}
Eq. \ref{12} can be more compactly expressed as

\begin{equation}\label{13}
\begin{split}
\frac{d F^S _2(x, t)}{d t} =  \alpha_s f(x,K)F^S _2(x, t) -\frac{\alpha_s ^2}{e^t}g(x,K)[F_2^S(x,t)]^2 = f^{(0)}F^S _2(x, t) -\frac{g^{(0)}}{e^t}[F_2^S(x,t)]^2,
\end{split}
\end{equation}

where $g(x,K)=\frac{5 K^2(x)}{6912\pi R^2}\cdot \left(2^{2\lambda +3}-1 \right)x^2 \xi (\lambda),\,\,f^{(0)}= \alpha_s \cdot f(x,K) ,\,\,g^{(0)}= \alpha_s ^2 \cdot g(x,K)$\\

\noindent The Eq. \ref{13} can be solved appropriately using Integrating Factor (IF). The solution has the following form:

\begin{equation}\label{14}
F_2 ^S (x,t)= \frac{e^{f^{(0)} t}(f^{(0)}-1)}{e^{(f^{(0)-1})t}g^{(0)} -C + f^{(0)}C},
\end{equation}
where $C$ is a constant to be determined using initial conditions. The value of $C$ will be different for both $x$ and $Q^2$ evolutions. To obtain the $x$ evolution of the singlet structure function at a particular $Q^2$, we will take an input at a larger value of $x$ and then we evolve the solution for all smaller values of $x$. Since, the proton's $F_2$ structure function is related to the singlet structure function $F_2 ^S$ (see Eq. \ref{2}) therefore, from now onwards we will drop the superscript ``S". So, the solution of $x$ evolution of proton's $F_2$ structure function comes out to be
\begin{equation}\label{15}
F_2  (x,t)= \frac{F_2 (x_0,t)\cdot (f^{(0)}-1)(f_0 ^{(0)}-1)e^{f^{(0)}t}}{e^{f_0 ^{(0)}t}(f^{(0)}-1)(f_0 ^{(0)}-1)+F_2 (x_0,t)\left[g^{(0)}(f_0 ^{(0)}-1)e^{(f^{(0)}-1)t} - g_0 ^{(0)}(f ^{(0)}-1)e^{(f_0 ^{(0)}-1)t} \right]},
\end{equation}
where $f_0 ^{(0)}= f^{(0)} (x_0)$, $g_0 ^{(0)}= g^{(0)} (x_0)$ and $F_2 (x_0,t)$ is the input at a larger $x_0$ value for a given value of $Q^2$ (or $t$). In the similar fashion, we can also work out for the $t$ (or $Q^2$) evolution of $F_2$. In this case, we will consider the input ( $F_2 (x,t_0)$ ) at a larger value of $Q^2$ for a particular $x$. The $Q^2$ evolution thus expressed as

\begin{equation}\label{16}
F_2  (x,t)= \frac{F_2 (x,t_0)\cdot (f^{(0)}-1)e^{f^{(0)}t}}{e^{f ^{(0)}t_0}(f^{(0)}-1)+F_2 (x,t_0)\cdot g^{(0)}\left[e^{(f^{(0)}-1)t} - e^{(f^{(0)}-1)t_0} \right]}
\end{equation}

\subsection{$\alpha_s$ as running coupling constant}

It is also important to include the $Q^2$ dependency of $\alpha_s$ into the GLR-MQ-ZRS equation. The running coupling $\alpha_s$ at leading order of QCD in terms of the variable $t$ has the following form

\begin{equation*}
\alpha_s (t) = \frac{4\pi}{\beta_0 t},\,\text{where}\, \beta_0 = 11-\frac{2}{3}n_f
\end{equation*}

Eq. \ref{13} can now be expressed as

\begin{equation}\label{17}
\frac{dF_2 ^S (x,t)}{dt}= \frac{1}{t}\tilde{f}(x,K)F_2 ^S (x,t)- \frac{1}{t^2 e^t}\tilde{g}(x,K)[F_2 ^S (x,K)]^2
\end{equation}
for which the solution is obtained using Integrating Factor (IF) as 

\begin{equation}\label{18}
F_2 ^S (x,t)= \frac{t^{\tilde{f}(x,K)}}{C-\tilde{g}(x,K)\Gamma[\tilde{f}(x,K)-1,t]},\,\,\text{where } \tilde{f}(x,K)=\frac{4\pi f(x,K)}{\beta_0}\,\,\text{and}\,\,\tilde{g}(x,K)=\frac{16\pi ^2g(x,K)}{\beta_0 ^2 \Lambda ^2}.
\end{equation}

Following the same procedure as in the case of constant $\alpha_s$, the $x$ and $Q^2$ (or $t$) evolutions of proton's $F_2$ structure function for $\alpha_s$ running are obtained separately and are given as follows:

x evolution of $F_2$:
\begin{equation}\label{19}
F_2 ^S (x,t)= \frac{F_2  (x_0,t)\cdot t^{\tilde{f}(x,K)}}{t^{\tilde{f}(x_0,K)}-F_2  (x_0,t)\left[\tilde{g}(x_0,K)\cdot\Gamma[\tilde{f}(x_0,K)-1,t]-\tilde{g}(x,K)\cdot\Gamma[\tilde{f}(x,K)-1,t]\right]}
\end{equation}

$Q^2$ (or $t$) evolution of $F_2$:
\begin{equation}\label{20}
F_2 ^S (x,t)= \frac{F_2  (x,t_0)\cdot t^{\tilde{f}(x,K)}}{t_0 ^{\tilde{f}(x_0,K)}-F_2  (x,t_0)\cdot\tilde{g}(x,K)\left[\Gamma[\tilde{f}(x,K)-1,t_0]-\Gamma[\tilde{f}(x,K)-1,t]\right]}
\end{equation}
Eq. \ref{19} and Eq. \ref{20} are expressed in terms of the incomplete gamma functions defined as $$\Gamma[a,x]=\int_{x}^{\infty} x^{a-1}e^{-x}dx.$$

On the other hand, the reduced neutral current (NC) differential cross section for $e^{+}p$ scattering after correcting for QED radiative effects is given by,
\begin{equation}\label{21}
\tilde{\sigma}_{NC} (x,Q^2,y)\equiv \frac{d^2\tilde{\sigma}_{NC} }{dx dQ^2}\cdot\frac{xQ^4}{2\pi\alpha_s ^2 Y_+}\equiv F_2 - \frac{y^2}{Y_+}F_L-\frac{Y_-}{Y_+}xF_3
\end{equation}

where
$Y_{\pm}= 1 \pm (1-y)^2 $, $y=\frac{Q^2}{sx}$ is the inelasticity and $s$ is the centre of mass squared energy of electrons and protons respectively.
For $Q^2 \leq 800 \,GeV^2$, the contribution of $Z$ exchange and the influence of $xF_3$ is expected to be small\cite{Andreev2}. So, $\tilde{\sigma}_{NC}\equiv  F_2 - \frac{y^2}{Y_+}F_L$. The longitudinal structure function $F_L(x,Q^2)$ is related to the longitudinal virtual photon absorption cross section, $\sigma_L$. The contribution of $F_L$ to the reduced cross section becomes significant at high value of inelasticity $y$. In the first approximation of quark parton model, $F_L$ is considered to be zero but in actual DIS experiments $F_L$ should be nonzero since it arises from gluon corrections. In the dipole model strict bound on the ratio of the structure functions $\frac{F_L}{F_2}$\cite{ewerz2008range} is provided as follows:
\begin{equation}\label{22}
\frac{F_L (x,Q^2)}{F_2 (x,Q^2)} \leq 0.27 \,
\end{equation}

It is also useful to define the cross section ratio $R$ of longitudinally to transversely polarised virtual photons which is related to
the structure functions $F_2$ and $F_L$ as

\begin{equation}\label{23}
R= \frac{\sigma_L}{\sigma_T}\equiv \frac{F_L}{F_2-F_L}
\end{equation}
Analysis of HERA data for $R(Q^2)$ averaged over $x$ for a wide range of $Q^2$ is found to be almost constant\cite{Andreev2}. 
Now, the expression of the longitudinal structure function $F_L(x,Q^2)$ is given by

\begin{figure}[t]
	\begin{subfigure}{.45\linewidth}
		\centering
		\includegraphics[width=.9\linewidth]{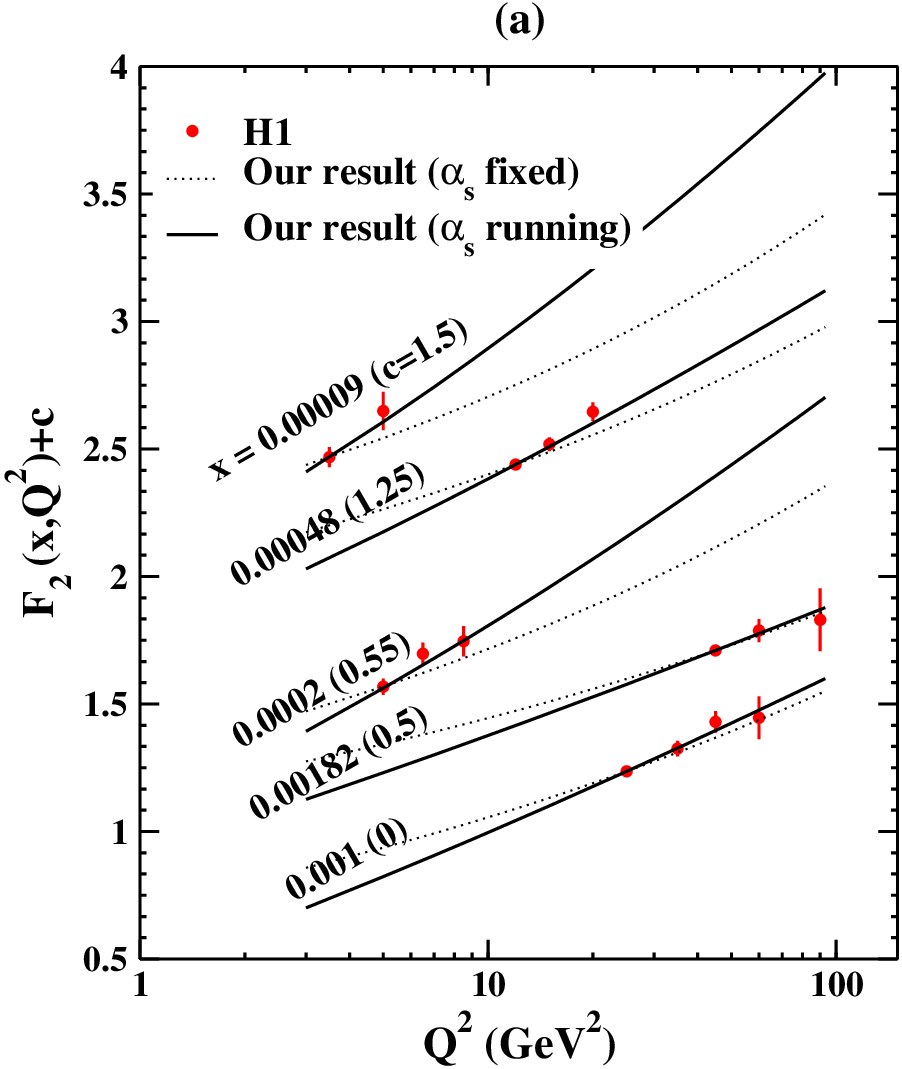}
		\label{fig:sub1}
	\end{subfigure}\\[5ex]
	\begin{subfigure}{.45\linewidth}
		\centering
		\includegraphics[width=.9\linewidth]{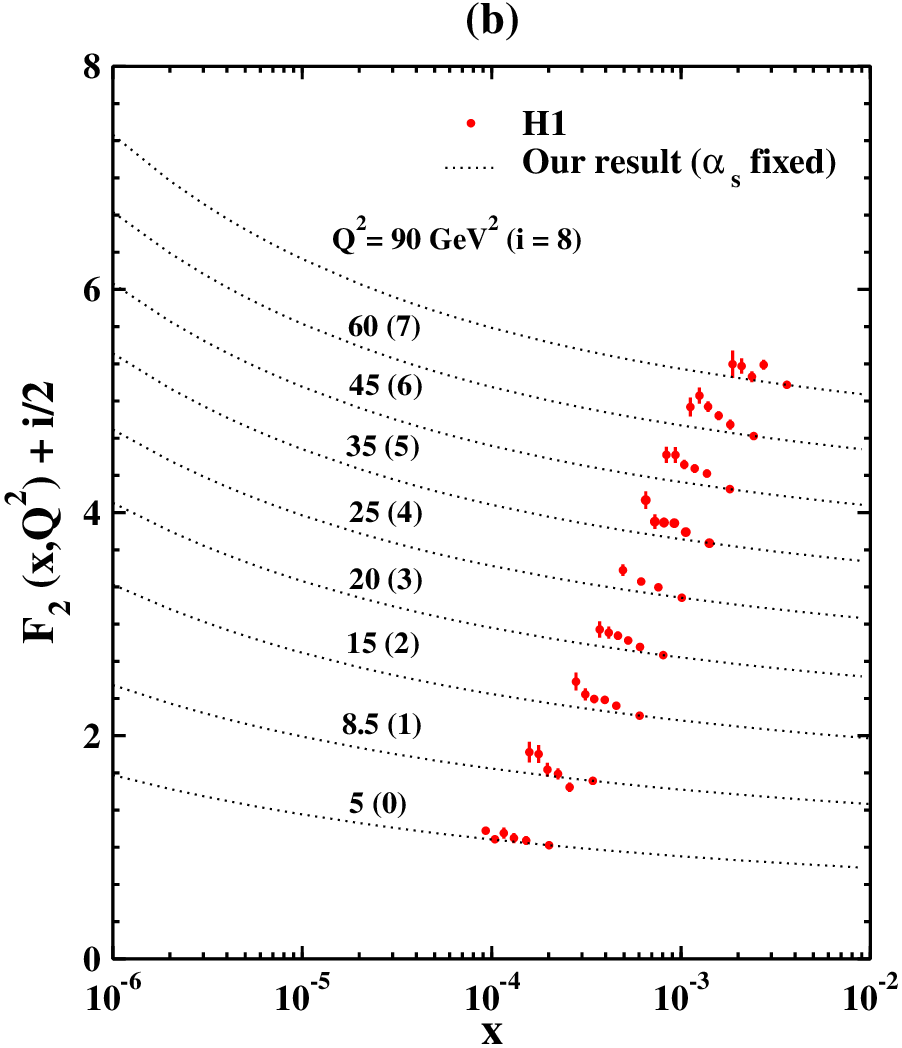}	
		\label{fig:sub2}
	\end{subfigure}\hfill
	\begin{subfigure}{.45\linewidth}
		\centering
		\includegraphics[width=.9\linewidth]{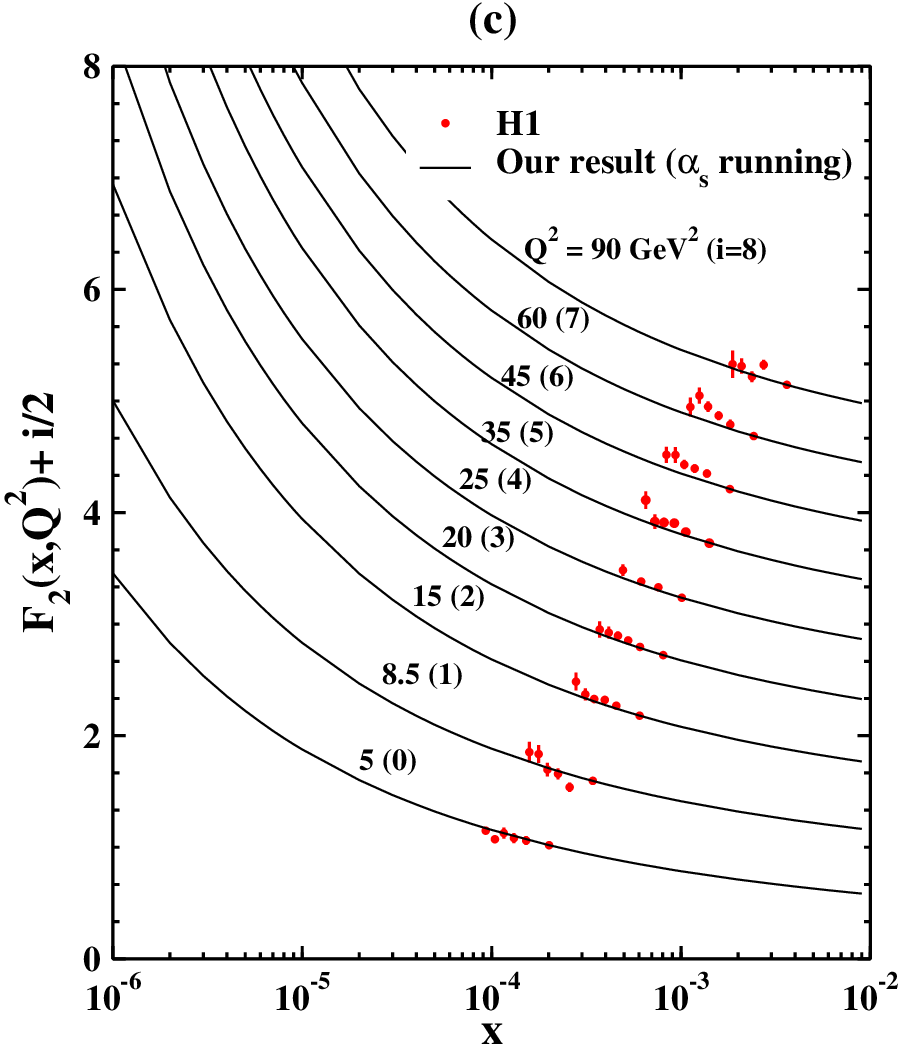}	
		\label{fig:sub3}
	\end{subfigure}
	\caption{\small In Fig. (a), $Q^2$ evolution of $F_2 (x,Q^2)$ for different values of $x$ is shown; the dotted lines represent our result at $\alpha_s$ fixed($=0.2$) and the solid lines represent our result running coupling $\alpha_s(Q^2)$. In Fig. (b) and (c), the $x$ evolution of $F_2 (x,Q^2)$ is shown for different values of $Q^2$. Our results are compared with the H1 data\cite{Andreev2}, the error bars represent total uncertainties on $F_2$. }
	\label{fig:test}
\end{figure}

\begin{equation}\label{24}
\begin{split}
F_L (x,Q^2)=& K^{S+NS}\otimes F_2 (x,Q^2) + K^G \otimes G(x,Q^2)\\
=& (K^{S+NS} + K^G)\cdot K(x)\otimes F_2 (x,Q^2),
\end{split}
\end{equation}
where both quarks and gluons contribute. Here the kernels $K^i$ are the coefficient functions and the convolution $\otimes$ denotes the usual prescription. The leading order coefficient functions for longitudinal structure functions can be found in \cite{MOCH2005123}. Since, we have already obtained the evolution of $F_2 (x,Q^2)$ hence the evolution of $F_L (x,Q^2)$ can be obtained using Eq. \ref{24}.

\begin{figure}[t]
	\centering
	\includegraphics[width=0.9\textwidth]{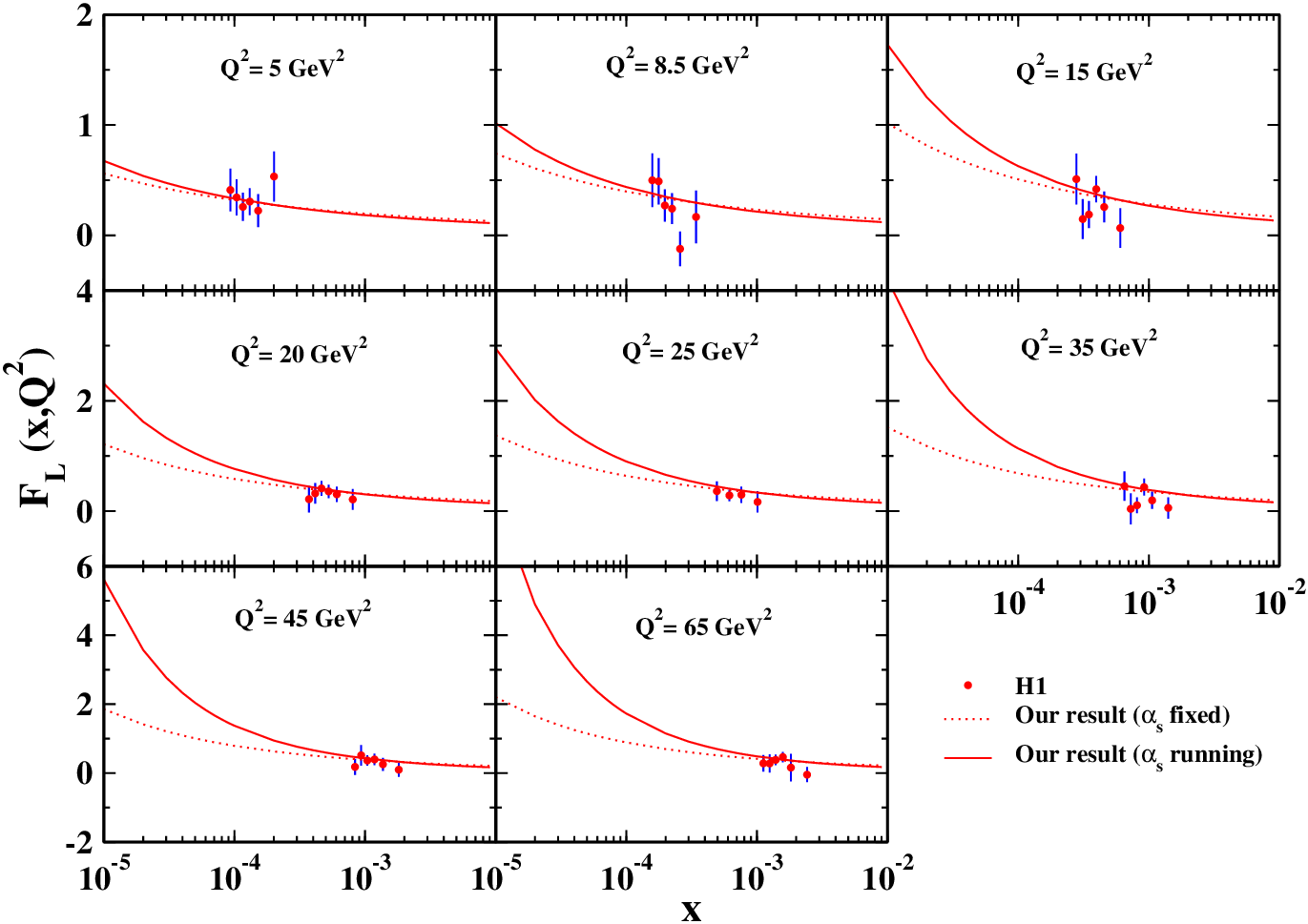}
	\caption{\small Our predicted results for the $x$ evolution of the longitudinal structure function $F_L (x,Q^2)$ for different $Q^2$ values is shown. The dotted lines represent our prediction at fixed $\alpha_s$ ($=0.2$) and the solid lines represent our predicted results for running coupling $\alpha_s (Q^2)$. Our results are compared with H1 data\cite{Andreev2}, the error bars represent total uncertainties on $F_L$.}
	\label{fig:1}       
\end{figure}

\section{Results and Discussion}\label{sec3}

In this paper, we have obtained analytical solutions of the DIS structure functions $F_2 (x,Q^2)$ and $F_L (x,Q^2)$ in the framework of GLR-MQ-ZRS equation, and the ratios $F_L (x,Q^2)/F_2 (x,Q^2)$ and $R(Q^2)=\sigma_L/\sigma_T$ are analysed in the kinematic range of $10^{-6}\leq x \leq 10^{-2}$ and $5\leq Q^2 \leq 100\,GeV^2$. While solving the GLR-MQ-ZRS equation for small-x region, we relied on Regge like behaviour of gluons. We have also assumed a direct relationship between the gluon distribution function and proton's $F_2$ structure function as stated in Eq. \ref{5}, the value of the parameters $a$ and $b$ in our analysis of the HERA data is found to be in the range of $1.2\,\text{to}\,1.7$ and $-0.11\,\text{to}\,-0.15$ respectively. It is to mention that in spite of the fact that GLR-MQ-ZRS equation was formulated two decades ago, very less work has been done in regards to phenomenological application of this equation. Some of the noted work on this equation can be found in these references \cite{zhu2016chaotic,zhu2016determination,zhu2}.

Our results of proton's $F_2$ structure function are shown in Fig. 1. We have compared our results with the high precision HERA H1 data\cite{Andreev2} as shown in the figure. In Fig. 1(a), the $Q^2$ evolution of $F_2 (x,Q^2)$ is presented for both the cases (i) $\alpha_s$ fixed at 0.2 (represented by the dotted lines) and (ii) $\alpha_s \rightarrow \alpha_s(Q^2)$ (represented by the solid lines). It can be seen from this graph that in case of running $\alpha_s$ our predicted results describe the available data quite well in a broad range of $x$ and $Q^2$, while, in case of fixed $\alpha_s$, our predicted results fail to agree with available H1 data in the region $x\leq 0.002$. The $x$ evolution of $F(x,Q^2)$ at different $Q^2$ values is presented in Fig. 1(b) for fixed $\alpha_s$, and for the running coupling $\alpha_s$ the $x$ evolution is plotted in Fig. 1(c). Clearly, our predicted results show better description of the H1 data\cite{Andreev2} when treating $\alpha_s$ to be free rather than a constant.

\begin{figure}[t]
	\centering
	\label{ fig7} 
	\begin{minipage}[t]{0.455\textwidth}
		\centering
		\includegraphics[width=.92\linewidth]{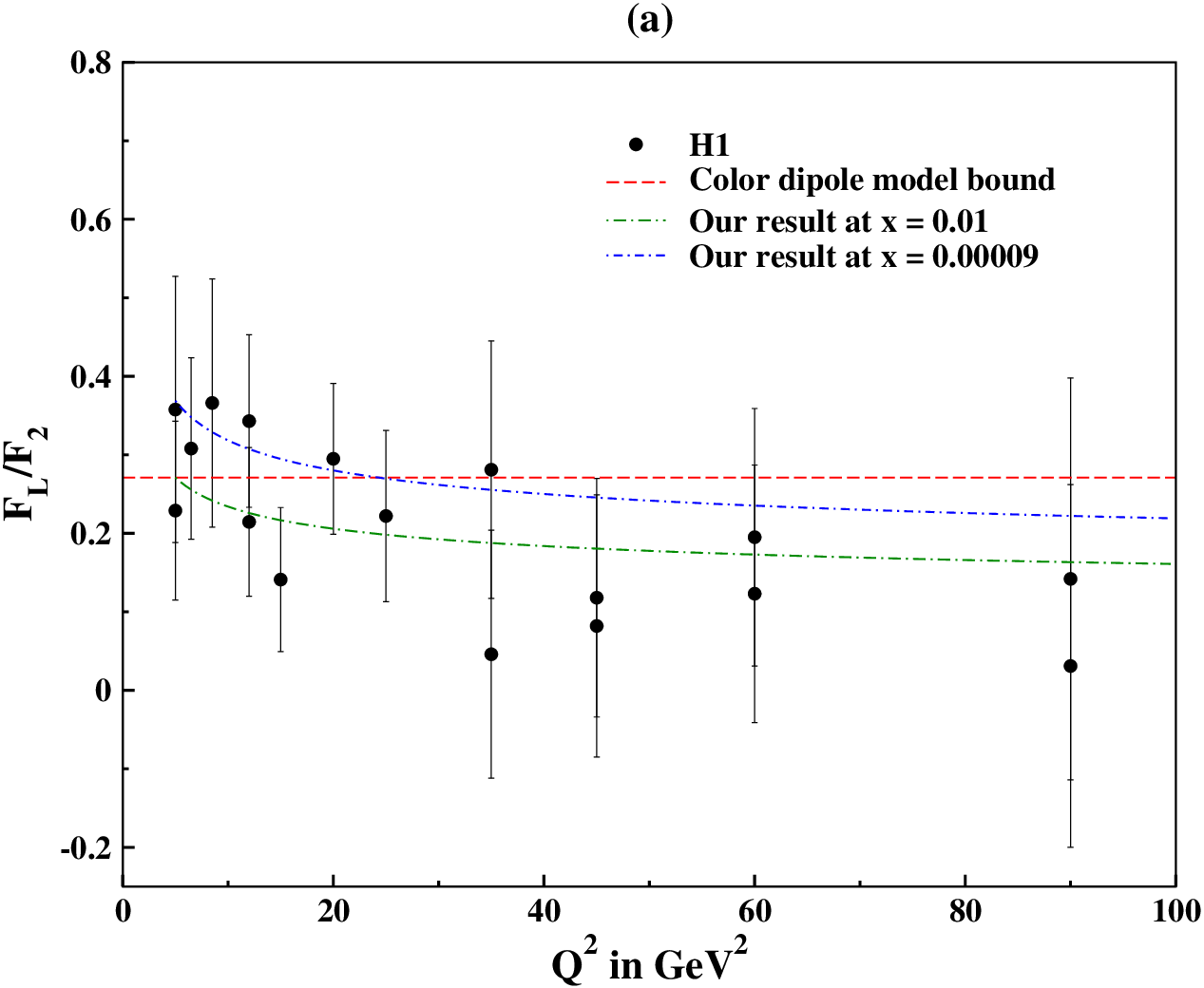} 
		
		
	\end{minipage}
	\hfill
	\begin{minipage}[t]{0.4\textwidth}
		\centering
		\includegraphics[width=.96\linewidth]{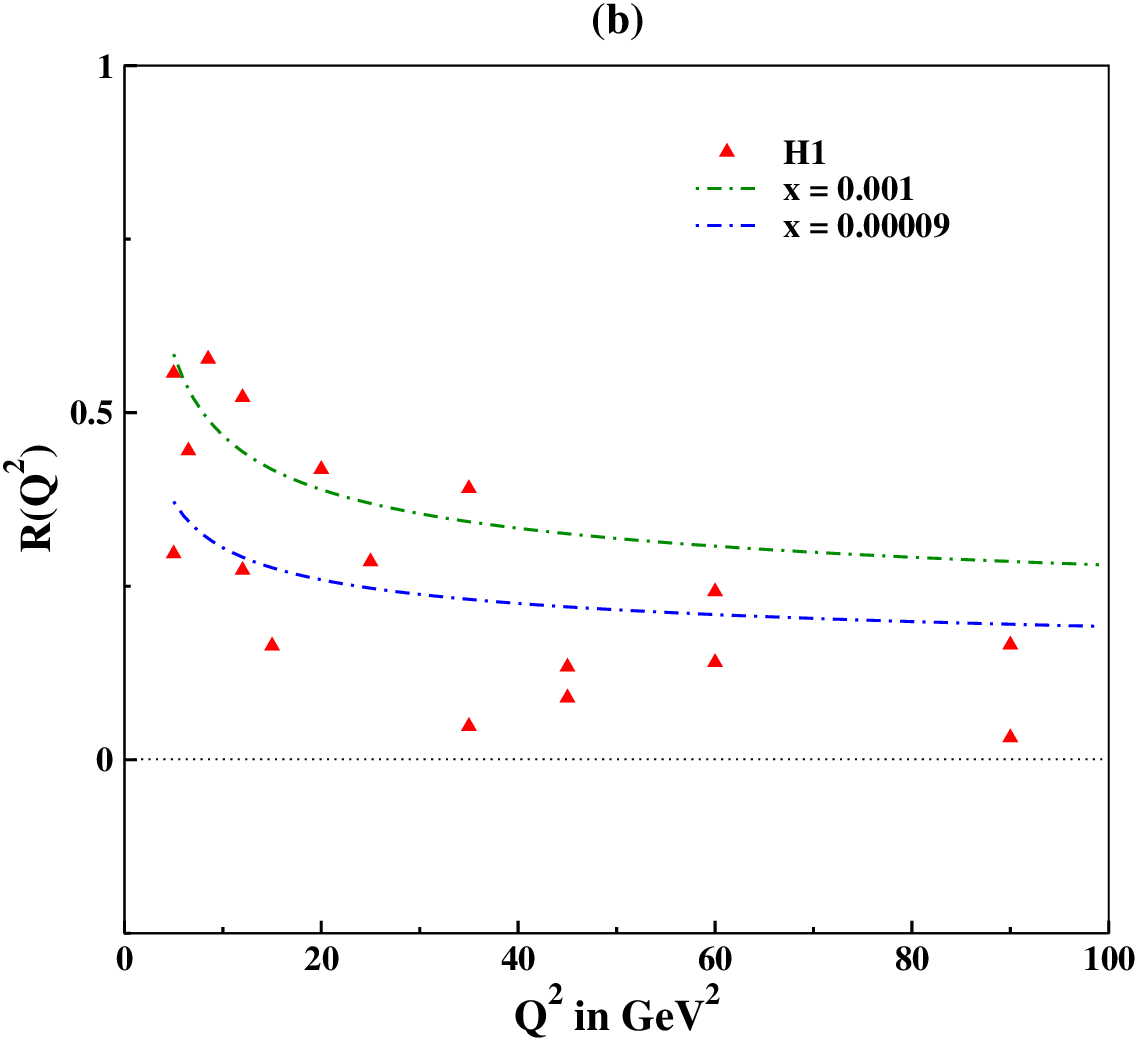} 
	\end{minipage}
	
	\caption{\small In figure (a) the ratio $F_L/F_2$ is plotted against $Q^2$ for two different values of $x= 10^{-2}\,\text{and}\,9\times10^{-5}$ respectively. The red dashed lines represent the bound given by colour dipole model\cite{EWERZ2007279,ewerz2008range}. Results are compared with H1 data\cite{Andreev2}, the error bars represent total error added in quadrature. In figure (b), the cross section ratio $R(Q^2)$ is plotted with respect to $Q^2$ for two different values of $x= 10^{-2}\,\text{and}\,9\times10^{-5}$ respectively. Our predicted results are compared with HERA H1 data.}
\end{figure}

In Fig. (2), our predicted results for the longitudinal structure function $F_L (x,Q^2)$ with respect to $x$ at different values of $Q^2$ is presented. Deviation of the results of $F_L (x,Q^2)$ at $\alpha_s$ fixed is observed from the results of $F_L (x,Q^2)$ for $\alpha_s (Q^2)$ towards small-x for a given $Q^2$, this deviation becomes more as $Q^2$ is increased. However, our predicted results agree reasonably well with the available experimental data\cite{Andreev2} in the given kinematic region.

In Fig. 3(a), the ratio $F_L (x,Q^2)/F_2 (x,Q^2)$ is plotted with respect to $Q^2$ at $x=0.01\,\text{and}\,0.00009$, $\alpha_s$ is considered to be free in this case. Our predicted results are compared with the H1 data\cite{Andreev2} and with the result obtained by the colour dipole model bound\cite{ewerz2008range}. The error bars of the ratio $F_L/F_2$ are determined by the following equation 

\begin{equation}\label{25}
\Delta(\frac{F_L}{F_2}) = \frac{F_L}{F_2}\sqrt{(\frac{\Delta F_L}{F_L})^2 + (\frac{\Delta F_2}{F_2})^2},
\end{equation}

\noindent where $\Delta F_L$ and $\Delta F_2$ are collected from the H1 experimental data in Ref. \cite{Andreev2}. Our predicted result of $F_L/F_2$ for a given $x$ almost remains constant for a broad range of $Q^2$, the ratio comes closer to the dipole bound at $x=0.00009$. The H1 data we have plotted in this figure have not been averaged over $x$. Finally, the cross section ratio $R(Q^2)$ is plotted in Fig. 3(b) for two different values of $x=0.001\,\text{and}\,0.00009$ respectively. The result of $R(Q^2)$ for a given $x$ almost remains steady in a broad range of $Q^2$, which is consistent with H1 results. The analysis of H1 data\cite{Andreev2} shows a constant value of $R = 0.23\pm 0.04$ over entire $Q^2$ region and  averaged over $x$. However, in this figure we have plotted the H1 data which are not averaged over $x$.   

\section{Conclusion}\label{sec4} 

In this paper we have obtained analytical solutions of the DIS structure function $F_2  (x,Q^2)$ using Regge like behaviour of gluons in the framework of GLR-MQ-ZRS equation. The longitudinal structure function $F_L (x,Q^2)$ is also obtained and studied using the solutions of $F_2 (x,Q^2)$. We have compared our results with HERA H1 experimental data and our predicted results are in good agreement. The accuracy of our results to describe the available experimental data enhances as we consider the strong coupling constant $\alpha_s$ to be free i.e. $\alpha_s (Q^2)$ rather than treating it as a constant. Our solution also gives a good description of the ratio $F_L/F_2$, our results at smaller-x
are close to the bound given by the dipole model approach. Also, our obtained results for the cross section ratio $R(Q^2)$ are consistent with the HERA H1 data. Finally, we conclude that the GLR-MQ-ZRS equation can be applied well in description of the experimental data and it can serve as an alternate equation to analyse experimental data at small-x. This work will encourage the researchers working in the phenomenology of QCD and we hope that this equation will help us understanding and exploring more about small-x physics in the coming years. 
\section*{Acknowledgement}

\noindent Madhurjya Lalung acknowledges Council of Scientific and Industrial Research, New Delhi for providing financial assistanship in the form of Senior Research Fellowship(SRF).
\bibliography{biblio.bib}
\end{document}